# Moduli Spaces of Phylogenetic Trees Describing Tumor Evolutionary Patterns

Sakellarios Zairis[1], Hossein Khiabanian[1], Andrew J. Blumberg[2], and Raul Rabadan[1]

[1] Department of Systems Biology, Columbia University
[2] Department of Mathematics, UT Austin

October 3, 2014

## Abstract

Cancers follow a clonal Darwinian evolution, with fitter subclones replacing more quiescent cells, ultimately giving rise to macroscopic disease. High-throughput genomics provides the opportunity to investigate these processes and determine specific genetic alterations driving disease progression. Genomic sampling of a patient's cancer provides a molecular history, represented by a phylogenetic tree. Cohorts of patients represent a forest of related phylogenetic structures. To extract clinically relevant information, one must represent and statistically compare these collections of trees. We propose a framework based on an application of the work by Billera, Holmes and Vogtmann on phylogenetic tree spaces to the case of unrooted trees of intra-individual cancer tissue samples. We observe that these tree spaces are globally nonpositively curved, allowing for statistical inference on populations of patient histories. A projective tree space is introduced, permitting visualizations of aggregate evolutionary behavior. Published data from three types of human malignancies are explored within our framework.

## 1 Introduction

A tumor is the result of successive accumulation of genetic alterations. As alterations accumulate, cancer cells of higher fitness replace earlier populations and drive progression of the disease. Senescence, apoptosis, immune surveillance, and drug treatment all represent selection pressures in the evolutionary environment of tumor cells. Competitive survival among tumor cells with periods of clonal replacement is a unifying characteristic of cancers, which are an otherwise diverse set of diseases. Darwinian evolution of tumor populations explains many important observations such as dynamic allele fractions of mutations and acquired resistance to chemotherapies. [9] A seminal paper by Peter Nowell in 1976 [15] first used the term "tumor stemlines" in its thesis that cancers derive from a common progenitor and progress via sequential selection among subclonal populations. It is increasingly being appreciated that the patterns of clonal evolution can vary widely between cancer types and between treatment strategies. Learning how these processes develop and how alterations accumulate provides valuable information in understanding the molecular mechanisms of oncogenesis and acquired drug resistance. A nuanced picture of an individual's tumor evolution may prove useful in predicting prognosis and tailoring disease management.

In the last five years, high throughput sequencing has illuminated the landscape of genomic alterations in a large number of tumors. The Cancer Genome Atlas (TCGA) and the International Cancer Genome Consortium (ICGC) have led this effort, sequencing thousands of tumors spanning the spectrum of human malignancies, leading to the identification of recurrent alterations that indicate mechanisms of convergent evolution in certain genes and pathways. Large cross-sectional studies, however, are not designed to capture the dynamic nature of tumor evolution. For this one needs longitudinal studies, sequencing a tumor at multiple time points within an individual's disease course, and such studies are only now emerging in the literature [7, 17, 19, 21]. Key questions in tumor dynamics surround the mechanisms of acquired drug resistance, the emergence of subclones with metastatic potential, the clinical stratification of patients according to observed tumor evolution, and the design of personalized drug treatment regimens to steer tumor evolution.

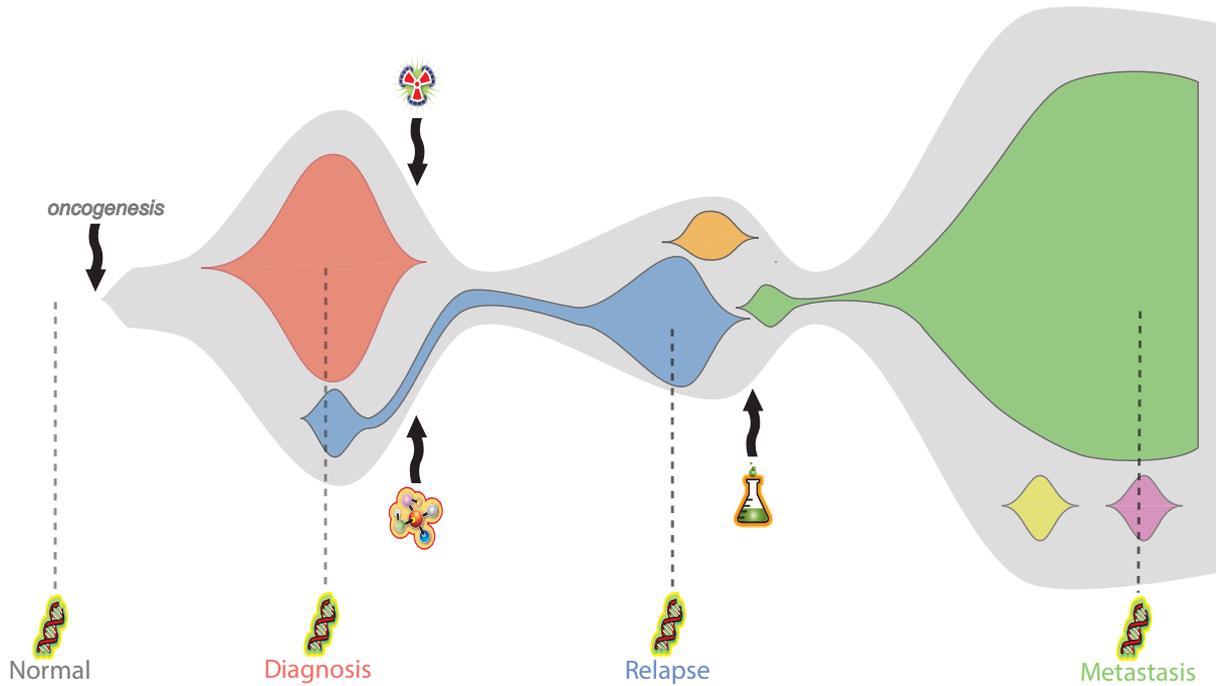

Figure 1: The dynamic nature of clonal evolution in cancer. We depict the evolution of a tumor through various clinical stages, with time running from left to right. There is an expansion of malignant cells beginning at "oncogenesis" and within the larger gray cell mass are contained different subclonal populations, represented in different colors. The overall size of the malignant cell mass is affected by therapeutic interventions, here depicted via symbols for radiation treatment, targeted molecular agents, and salvage chemotherapy. The schematic shows a progression from oncogenesis through clinically distinct phases: primary tumor, remission following initial therapy, relapse, remission following salvage therapy, and finally uncontrolled metastatic spread. Sequencing may be performed at multiple clinical time points, however the mutational spectrum primarily reflects the dominant clone at that time point.

The rise of dominant clones in a tumor can be inferred by sequential sequencing of an individual's disease. Successive genomic snapshots may represent defined temporal intervals, progression of disease through predefined stages, or successive anatomic sites to which a cancer has spread. In all of these scenarios we make observations about an evolutionary process and can represent the relationships between genomic snapshots as a phylogenetic tree. When large cohorts of cancer patients are studied, yielding a forest of such phylogenetic trees, we require a mathematical framework in which to reason about aggregate evolutionary behaviors. We must be able to, at a minimum, directly compare two evolutionary histories. We would like to be able to define distances between trees, stratify patients according to the evolutionary trajectories of their cancer, and correlate different trees with prognostic variables and molecular markers. A more ambitious aim would be to assess statistical significance on samples of trees and implement machine learning algorithms which act directly upon forests of phylgenetic structures. Here, we apply the work of Billera, Holmes and Vogtmann [4] and Sturm [20] on geometric spaces of phylogenetic trees to provide a framework for the visualization and quantitative summarization of tumor evolutionary histories. Our work provides a compact language for biologists and oncologists to use in describing longitudinal cancer genomic data sets.

We first describe the general space of tumor evolutionary histories and then elaborate on the most common cases of having either three or four samples per patient. Section 2 describes the topology and geometry of the tree spaces, along with the basic summary statistics they allow, and also introduces a projective tree space that will underlie our visualizations. It provides a theoretical background of the main concepts that will be applied in this and future work. Section 3 demonstrates the application of the projective moduli spaces in visualizing the aggregate evolutionary behavior of two hematologic malignancies, acute myeloid leukemia and follicular lymphoma, as well as one solid tumor, pancreatic adenocarcinoma. We conclude by discussing the exciting theoretical and applied frontiers that remain open in the development of evolutionary moduli spaces for longitudinal cancer genomic data.

## 2 Description of the space of trees

We can understand the relation between $m$ different genomes following clonal evolution as a phylogenetic tree with $m$ leaves. A phylogenetic tree is a weighted, connected graph with no circuits, having $m$ distinguished vertices of degree 1 labeled $\{1, \ldots, m\}$, and all other vertices of degree $\geq 3$. Edges that terminate in leaves are "external" edges and the remaining edges are "internal".

When the external branches all have length 0, the tree space we have described (where the nonzero weights are on the internal branches) was introduced and studied by Billera, Holmes, and Vogtmann. Specifically, the structure of the internal branches is captured by the $\text{BHV}_{m-1}$ construction (where the $m-1$ index arises from the fact that they consider rooted trees). Allowing potentially nonzero weights for the $m$ external leaves corresponds to crossing with an $m$-dimensional orthant. Therefore, the space we wish to study is simply

$$\Sigma_m = \text{BHV}_{m-1} \times (\mathbb{R}^{\geq 0})^m.$$

We refer to $\Sigma_m$ as the evolutionary moduli space.

### 2.1 The metric geometry of evolutionary moduli spaces

As described above, the space $\Sigma_m$ is just a set of points. The key insight of Billera, Holmes, and Vogtmann is that this space is equipped with a natural metric that endows the space with an intrinsic geometry.

The metric on $\text{BHV}_{m-1}$ is induced from the standard Euclidean distance on each of the orthants, as follows. For two trees $t_1$ and $t_2$ which are both in a given orthant, the distance $d_{\text{BHV}_{m-1}}(t_1, t_2)$ is defined to be the Euclidean distance between the points specified by the weights. For two trees which are in different quadrants, there exist (many) paths connecting them which consist of straight lines in each quadrant. The length of such a path is the sum of the lengths of these lines, and the distance $d_{\text{BHV}_{m-1}}(t_1, t_2)$ is then the minimum length over all such paths. There is an analogous metric on $\Sigma_m$, which can be regarded as induced from the metric on $\text{BHV}_{m-1}$. Specifically, for a tree $t$, let $t(i)$ denote the length of the external edge associated to the vertex $i$. Then

$$d_{\Sigma_m}(t_1, t_2) = \sqrt{d_{\mathrm{BHV}_{m-1}}(\bar{t}_1, \bar{t}_2) + \sum_{i=1}^{m}(t_1(i) - t_2(i))^2},$$

where $\bar{t}_i$ denotes the tree in $\mathrm{BHV}_{m-1}$ obtained by forgetting the lengths of the external edges (e.g., see [18]).

The metric space $(\Sigma_m, d_{\Sigma_m})$ allows us to talk meaningfully about the distance between two evolutionary histories. But more importantly, $d_{\Sigma_m}$ allows us to describe the geometry of $\Sigma_m$, specifically curvature. The curvature of a space can be seen in the behavior of triangles; given side lengths $(\ell_1, \ell_2, \ell_3) \subset \mathbb{R}^3$, a triangle with these side lengths on the surface of the Earth is "fatter" than the corresponding triangle on a Euclidean plane. We can be more precise about this by looking at the distance from a vertex of the triangle to a point $p$ on the opposite side — in a fat triangle, this distance will be larger than in the the corresponding Euclidean triangle. (Thin triangles are defined analogously.)

Alexandrov observed that this perspective makes sense in any geodesic metric space [2]. A metric space $M$ is a geodesic metric space if any two points $x$ and $y$ can be joined by a path with length precisely $d(x, y)$. Then given points $p, q, r$, we have the triangle $T = [p, q, r]$ with edges the paths connecting each pair of vertices. These paths specify edge lengths, and so we can find a corresponding triangle $\tilde{T}$ in Euclidean space. Given a point $z$ on the edge $[p, q]$, a comparison point in $\tilde{T}$ is a point $\tilde{z}$ on the corresponding edge $[\tilde{p}, \tilde{q}]$ such that $d(\tilde{z}, \tilde{p}) = d(p, z)$.

We say that a triangle $T$ in $M$ satisfies the CAT(0) inequality if for every pair of points $x$ and $y$ in $T$ and comparison points $\tilde{x}$ and $\tilde{y}$ on $\tilde{T}$, we have $d(x, y) \leq d(\tilde{x}, \tilde{y})$. If every triangle in $M$ satisfies the CAT(0) inequality then we say that $M$ is a CAT(0) space. More generally, let $M_\kappa$ denote the unique two-dimensional Riemannian manifold with curvature $\kappa$. Then we say that a geodesic metric space $M$ is CAT($\kappa$) if every triangle in $M$ satisfies the inequality above for the comparison triangle in $M_\kappa$. Gromov gave a condition for a cubical complex to be CAT(0), and using this condition Billera, Holmes, and Vogtmann showed that $\mathrm{BHV}_{m-1}$ is CAT(0). An immediate generalization of their argument yields the analogous result for $\Sigma_m$.

**Theorem 1.** *The space $\Sigma_m$ is a CAT(0) space.*

## 2.2 Statistics on evolutionary moduli spaces

Sequencing longitudinal cancer samples can be regarded as sampling from a distribution on the evolutionary moduli space. Differences between the distributions associated with different tumors can be used to predict different evolutionary trajectories. First, we must justify the use of distributions on $\Sigma_m$. One can set up many aspects of the formal apparatus of probability theory on any complete metric space with a countable dense subset (i.e., a Polish spaces).

**Theorem 2.** *The space $\Sigma_m$ is a Polish space.*

There are many natural distributions on $\Sigma_m$ (e.g., uniform selection of tree topology followed by uniform selection of edge weights from a range $[a, b]$), although constructing biologically meaningful distributions appears to be a challenging problem. The principal virtue of establishing that $\Sigma_m$ is a CAT(0) space is that in this context, there exist well-defined notions of mean and variance, and more generally one can attempt to perform statistical inference. Holmes has written extensively on this topic [10, 11, 12] in the specific context of tree spaces (and see also [8]), and an account of basic statistical procedures for CAT(0) spaces in general has been given by Sturm [20]. The correct notion of the mean of a set of points is a generalization of a centroid: we define the Fréchet mean and variance in $\Sigma_m$.

**Definition 1.** *Given a fixed set of $n$ trees $\{T_0, \ldots T_{n-1}\} \subseteq \Sigma_m$, the Fréchet mean $T$ is the unique tree that minimizes the quantity*

$$E = \sum_{i=0}^{n-1} d_{\Sigma_m}(T_i, T)^2.$$

*The variance of $T$ is the ratio $\frac{E}{n}$.*

Sturm's work provides an iterative procedure for computing the mean in $\Sigma_m$, and by exploiting the local geometric structure of $\Sigma_m$, Miller, Owen, and Provan produce somewhat more efficient algorithms for computing the mean.

There are many natural test statistics defined in terms of the Fréchet mean and candidate inference algorithms that use the metric structure of tree space. One can use resampling and Monte Carlo simulation to obtain confidence intervals and perform inference, but practical study of such procedures and the development of the theoretical foundations for inference are both active areas of inquiry (e.g., the recent work in the context of tree spaces [10, 11, 12, 3, 8]). In our forthcoming paper [22], we develop and apply such techniques to inference and machine-learning problems arising in the study of tumor evolution.

## 2.3 The projective evolutionary moduli spaces

We are primarily interested in classifying and comparing distinct evolutionary behaviors by understanding the relative lengths of edges: rescaling edge lengths does not change the relationship between the branches. Thus, we will also use the quotient space of $\Sigma_m$ by the equivalence relation that for each orthant, the tree $\{t_i\}$ is equivalent to the rescaled tree $\{\lambda t_i\}$, i.e., the subspace of $\Sigma_m$ consisting of the points $\{t_i\}$ in each orthant for which the constraint $\sum_i t_i = 1$ holds. We will denote this quotient by $\mathbb{P}\Sigma_m$, the evolutionary projective moduli space.

This space of trees (without external edges) of fixed length was studied by Boardman and is denoted by $\tau_{m-1}$. The space of $m$ external branches that sum to a fixed length is an $m-1$ dimensional simplex, which we denote $T_{m-1}$. As we are requiring that the length of internal branches plus the external branches sum to a fixed constant, we can describe our space as the join of $\tau_{m-1}$ and $T_{m-1}$:

$$\mathbb{P}\Sigma_m = \tau_{m-1} \star T_{m-1}.$$

Since for the applications we describe herein the trees have either 3 or 4 leaves, it is instructive to describe explicitly the spaces $\mathbb{P}\Sigma_3$ and $\mathbb{P}\Sigma_4$. In the case of 3 leaves all structure is in the external branches and $\mathbb{P}\Sigma_3$ is a triangle. The triangle has three vertices and 3 edges; below we provide biological interpretations for these regions of the space in the context of different experiments. In the case of four leaves, $\tau_3$ is a set of 3 points reflecting the three possible topologies of unrooted 4-trees and $\mathbb{P}\Sigma_4$ becomes a richer space in which to compare and visualize evolutionary modes.

There is a natural projection map $\Sigma_m \to \mathbb{P}\Sigma_m$ given by rescaling. However, a number of warnings apply to the use of this projection. Notably, $\mathbb{P}\Sigma_m$ is not a CAT(0) space (it is a CAT(1) space). As a consequence, we cannot compute meaningful averages or variances in general. Moreover, the metric structure on $\mathbb{P}\Sigma_m$ is complicated. Even for a single simplex $\Delta_n$, treating $\Delta_n$ as a subspace of $\mathbb{R}^n$ does not lead to sensible statistical procedures. In this case, an approach to inference was introduced by Aitchison [1]. In work in progress we are studying the integration of Aitchison's transformation with $\mathbb{P}\Sigma_m$.

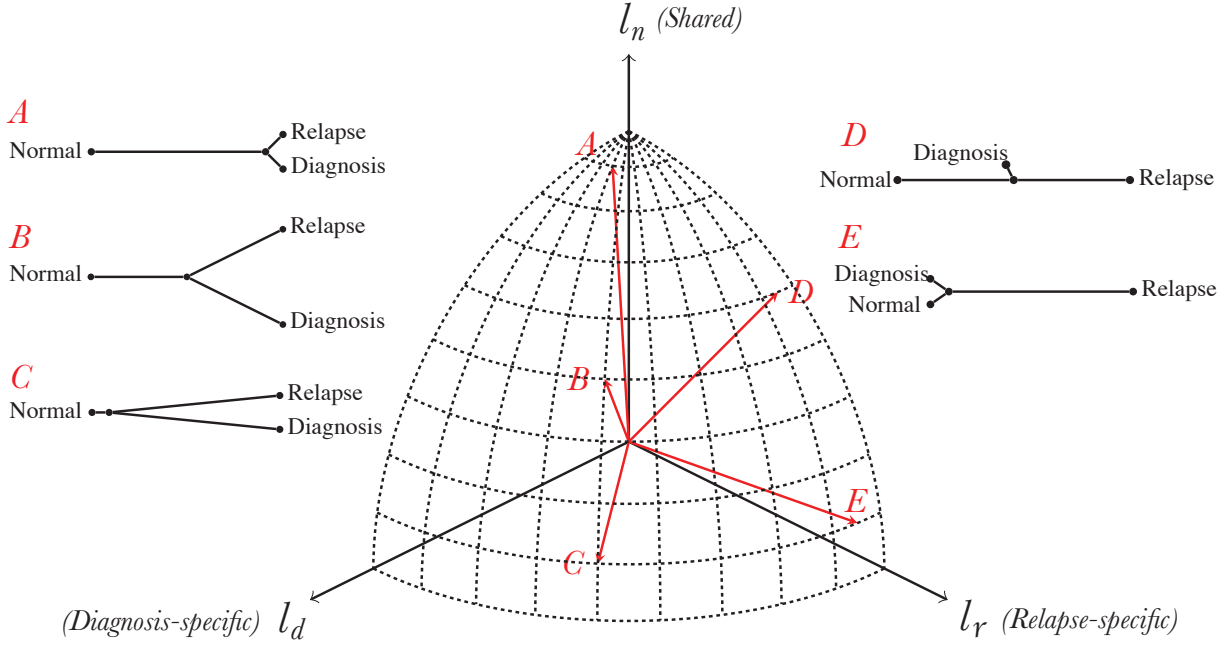

Figure 2: Evolutionary modes in $\mathbb{P}\Sigma_3$. A: frozen evolution, B: branched evolution, C: divergent evolution, D: linear evolution, E: somatic hypermutation.

## 2.4 Triplet data

In the context of cancer patients, triplet samples are often comprised of 1) **normal** tissue, 2) malignant tissue at **diagnosis**, and 3) malignant tissue at a later clinical time point such as local **relapse**. The moduli space of unrooted phylogenetic 3-trees, $\Sigma_3$, is a Euclidean 3-orthant whose basis vectors represent the 3 external edge lengths $(l_n, l_d, l_r)$. We project each tree onto $\mathbb{P}\Sigma_3$, the space formed by the intersection $\mathbb{R}^{3+} \cap S^2$, by rescaling the branch lengths. This space is visualized in Figure 2

The general case of three nonzero external branch lengths is called branched evolution and such phylogenetic trees will be found far from the boundary of $\mathbb{P}\Sigma_3$. We would also like to understand the possible singular cases that occur when one or more branches degenerate. If all branch lengths are zero then we have the trivial situation of no evolution among the three samples.

The edges of $\mathbb{P}\Sigma_3$ represent trees in which a single branch has collapsed to zero. As $l_n \to 0$ we have the situation where the diagnosis and the relapse are completely distinct tumors whose earliest common ancestor is in fact normal tissue. We call this divergent evolution. As $l_d \to 0$ we have the situation where the diagnosis is a perfect intermediate between the normal and relapse genotypes, the well known case of linear evolution. Lastly, as $l_r \to 0$ we have the situation where the relapse sample is actually the intermediate between normal and diagnosis genotypes, indicating the emergence of an ancient clone that was not dominant at the time of diagnosis. We call this revertant evolution.

The vertices of $\mathbb{P}\Sigma_3$ represent trees in which two branches have collapsed to zero. Near the "shared" vertex is the case where the tumor genomics are almost identical between diagnosis and relapse samples with respect to normal tissue. From a clinical perspective, no further mutations are needed beyond the diagnosis stage for the disease to relapse, and we term this scenario frozen evolution. Near the "diagnosis" vertex is the case where the relapsed tumor is almost identical to normal, healthy tissue with respect to the lesion at diagnosis. This would be a highly unusual set of genotypes to observe since advanced cancers require some genomic deviation from normal. Near the "relapse" vertex is the case where the tumor at diagnosis is essentially the same as normal tissue compared to the number of mutations specific to the relapsed disease. Rapid accumulation of mutations can result from a shifting fitness landscape during medical therapy, and this region of the space can indicate somatic hypermutation. This scenario does not imply that the lesion at diagnosis has zero difference from normal tissue, but rather that the difference is dwarfed by the number of mutations accumulated in the relapsed sample.

## 2.5 Quadruplet data

Quadruplet samples can arise from 1) **normal** tissue, 2) malignant tissue at **diagnosis**, and 3) malignant tissue at local **relapse** and 4) malignant tissue from distant **metastasis**. Unrooted trees constructed from quadruplet data contain a single internal edge, implying 3 possible tree topologies. We decompose the moduli space of unrooted phylogenetic 4-trees, $\Sigma_4$, into the product of spaces for its internal and external edges respectively, $BHV_3 \times \mathbb{R}^{4+}$. Upon rescaling of the branch lengths we project each tree onto $\mathbb{P}\Sigma_4$, the space formed by $\tau_3 \star T_3$, which is the join of a set of three points and a tetrahedron.

In Figure 3, we illustrate the two components of $\mathbb{P}\Sigma_4$. A quadruplet is represented by a point in the star plot and a point in the tetrahedron. The three arms of the star plot represent the three possible tree topologies, and they meet at an origin corresponding to the degenerate case of a length zero internal branch. The vertices of the tetrahedron correspond to trees having only one nonzero external branch, the edges to trees with two nonzero external branches, and the faces to trees with three nonzero external branches.

## 2.6 Quintuplet data

Quintuplet samples might include 1) **normal** tissue, 2) malignant tissue at **diagnosis**, 3) malignant tissue at local **relapse**, 4) malignant tissue from distant **metastasis**, and 5) malignant tissue collected at **autopsy**. Unrooted trees constructed from quintuplet data contain two internal edges, implying 15 possible topologies. We decompose the moduli space of unrooted phylogenetic 5-trees, $\Sigma_5$, into the product of spaces for its internal and external edges respectively, $BHV_4 \times \mathbb{R}^{5+}$. Upon rescaling of the branch lengths we project each tree onto $\mathbb{P}\Sigma_5$, the space formed by $\tau_4 \star T_4$. The internal space of $\Sigma_5$ can be thought of as a cone on $\tau_4$, the Petersen graph. This object is a cubic graph with no planar embedding whose shortest circuit is 5. In Figure 4 we have arranged the possible tree topologies along the 15 edges of $\tau_4$, and each vertex corresponds to an intermediate point of rotations between three adjacent topologies. We represented the space of external branches for triplet data as a 2–simplex, for quadruplet data as a 3–simplex, and naturally for quintuplet data the space will be a 4–simplex, also called a pentachoron. Explicit schemes for visualization of $\mathbb{P}\Sigma_5$ must encode the radial coordinate along the cone on $\tau_4$ and also provide lower dimensional projections of the data residing in the pentachoron.

From a visualization standpoint, the only impediment to explicit constructions of $\mathbb{P}\Sigma_m$ for $m \geq 6$ is the increasing dimensionality of the space. The number of possible topologies of an unrooted phylogenetic tree on $m$ samples grows as $(2m-5)!!$ and the overall dimension of $\Sigma_m$ is $2m-3$.

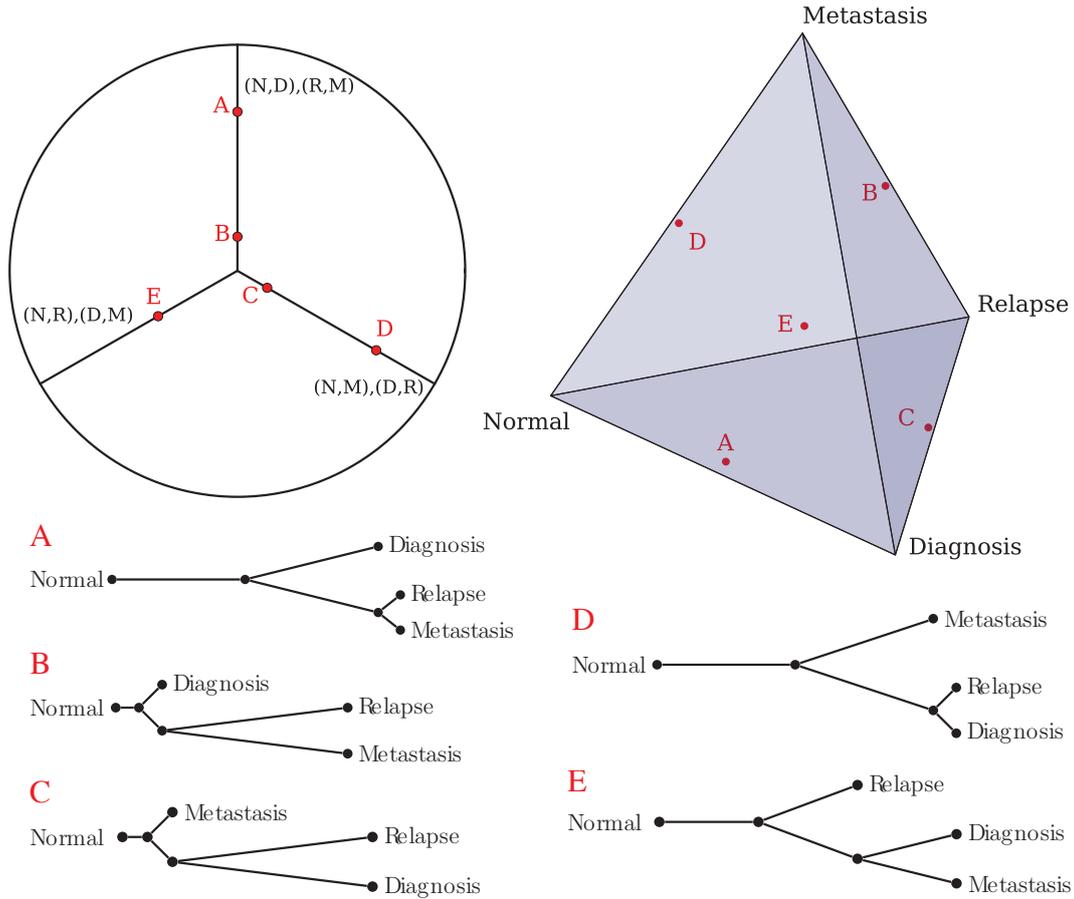

Figure 3: Evolutionary modes in $\mathbb{P}\Sigma_4$. Each tree, A—E, is represented by a pair of points. The three arms of the star plot specify the internal branch, and thus the topology, while the tetrahedral plot describes the external branches.

## 3 Case Studies

We now turn to the recent cancer genomics literature for examples of different patterns of tumor evolution. We examine the progression of two major hematologic malignancies for which there is abundant triplet data: acute myelogenous leukemia and follicular lymphoma. Then we shift to the highly aggressive solid tumors arising in the exocrine pancreas, pancreatic ductal adenocarcinoma, for which there is publicly available quadruplet data. In each case study we visualize aggregate evolutionary behavior and compute centroid trees via an implementation of Definition 1 (Section 2.2) given by [14]. Useful acronyms for the ensuing sections are WGS (whole genome sequencing) and WES (whole exome sequencing).

### 3.1 Relapsed acute myelogenous leukemia

Acute myelogenous keukemia (AML) accounts for about 80% of acute leukemia in adults with a median survival time of less than three years, accounting for more than 1% of cancer deaths in the US. AML is caused by the abnormal rapid growth of myelogenous progenitor cells interfering with normal hematopoeisis. Most patients die from relapse after chemotherapy and subsequent disease progression, with relapse free survival at only 40%. [16] The molecular mechanism of relapse in AML is not fully understood. In a recent study of relapsed AML, [7] the evolution to relapse was followed in 8 patients who received both induction and consolidation chemotherapy.

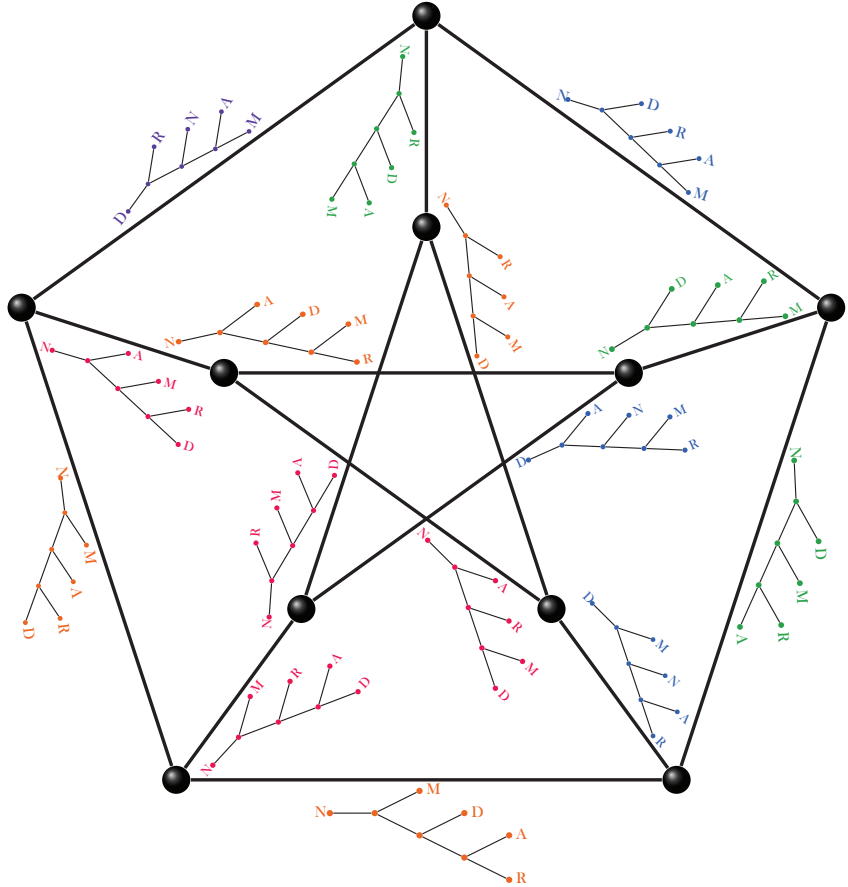

Figure 4: The 15 possible tree topologies arranged on the edges of $\tau_4$ and color coded to reflect biological plausibility. Warmer colors correspond to evolutionary relationships on the five samples that would be highly surprising, such as **normal** tissue and the tumor at **autopsy** being adjacent in the phylogeny.

The time to relapse varied between 235 - 961 days, and the investigators found that treatment did not eliminate the original cancer clone in any of the patients. WGS was performed on all 8 patients and an average of 21 protein changing mutations per patient were revealed.

When mapped to $\mathbb{P}\Sigma_3$ all patients are near the "frozen evolution" vertex illustrating that very few mutations were specific to either diagnosis or relapse samples (Figure 5). Despite the genotypes being virtually shared between diagnosis and relapse, the latter is a far more dangerous clinical entity. Also worth noting is the recurrence of mutations in genes that regulate DNA methylation such as *DNMT3A*, *IDH1*, and *IDH2*. The combination of frozen evolution, mutations that could affect global methylation patterns, and clinical progression of disease hints that the majority of evolution in this cancer is occurring beyond the DNA level. Indeed, recent reports suggest that relapse in AML is driven by epigenetic deregulation. [13]

## 3.2 Follicular lymphoma transformation to diffuse large B-cell lymphoma

Follicular lymphoma (FL) is a common lymphoid cancer, comprising 13% of all mature B-cell neoplasms. Roughly 20% of FL cases undergo a histologic transformation to more aggressive lymphoma phenotype resembling diffuse large B-cell lymphoma. While the prognosis of FL as a whole is 80% at 5 years, the prognosis for tFL is far worse at 20% survival after 2 years. Two recent genomic studies independently analyze patients with FL–tFL sample pairs and assess the evolutionary behavior of the transformation. In the first paper, [19] WES was performed on 12 patients, only 4 of whom had matched normal tissue. For the 8 patients in which somatic mutations could not be reliably called, a panel of 52 genes with established roles in lymphomagenesis was used as a surrogate genotype. In the second paper, [17] WES was performed on 4 patients and WGS was performed on 6 patients. We pool this data into three distinct groups based on genotype construction: WGS (6 patients), WES (8 patients), and curated gene panel (8 patients).

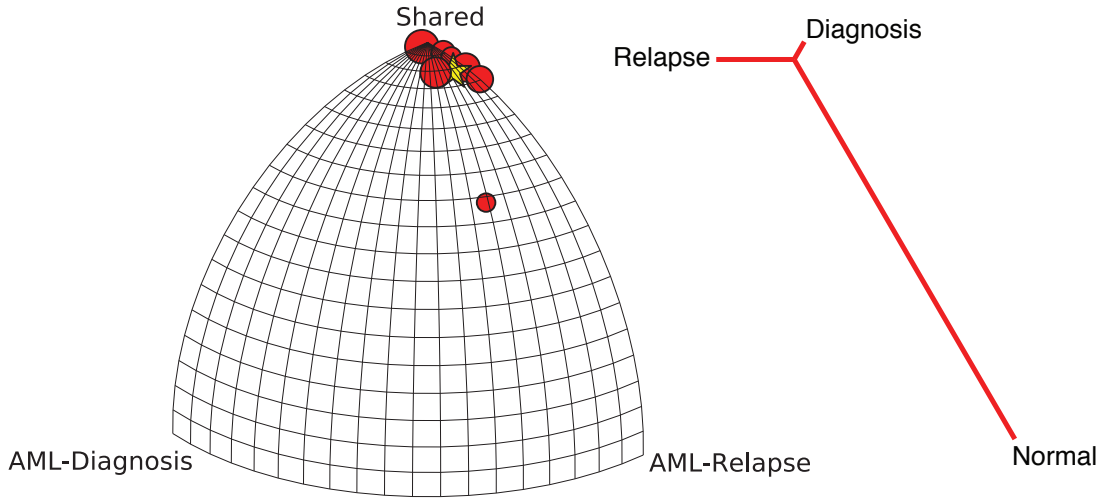

Figure 5: Frozen linear evolution from diagnosis to relapse observed in 8 AML patients. Patients are represented by red circles, scaled by their total number of mutations. The centroid of the distribution is represented as a gold star, and its associated phylogenetic tree is visualized.

The pooled data is visualized in Figure 6, with different colorings for the three groups. Until recently there was no consensus on the evolutionary mode of FL–tFL transformation, with some data even pointing to a linear process.[6] Figure 5, however, clearly demonstrates that the majority of the data fall in the bulk of $\mathbb{P}\Sigma_3$ and represent a branched evolutionary process. Comparison between the different genotype constructions further reveals that the degree of branched vs. linear evolution observed depends on the data curation strategy.

### 3.3 Metastatic pancreatic cancer

Cancer of the exocrine pancreas accounts for roughly 85% of all pancreatic malignancies and is the 4th leading cause of cancer-related deaths in the United States. In a recent study, [5] 13 cases of widely metastatic pancreatic ductal adenocarcinoma (PDAC) were studied at autopsy using a genome-wide detection method of structural rearrangements. The anatomic sites represented among the metastases include liver, lung, diaphragm, adrenal glands, peritoneum, and omentum.

We are interested in evolutionary histories involving distinct anatomical regions. To cast this data as quadruplets of successive anatomic sites of disease, we consider the hypothesis that the liver should represent the first metastatic location of PDAC. There is direct anatomical communication between the exocrine pancreas and the liver, via the common bile duct, while many of the other metastatic sites are only reachable via hematogenous spread of cancer cells. Furthermore, the liver receives a large fraction of cardiac output and might therefore be responsible for seeding the various more distant sites via the blood. For these reasons we are interested in differentiating between metastases to the liver vs. other sites. We partition the large number of samples per patient into the following disjoint subsets: normal tissue (1), primary pancreatic

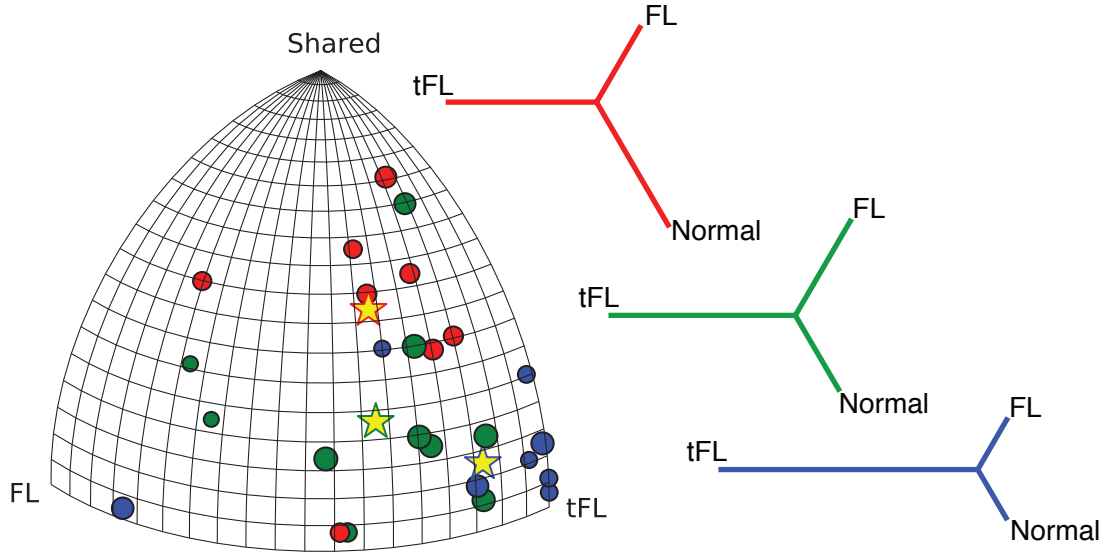

Figure 6: Different degrees of branched evolution observed in patients with FL–tFL transformation. Colored circles are scaled by patients' total number of mutations. Genotype construction strategy affects the degree of branched vs. linear evolution observed in WGS (red), WES (green), and curated gene panel (blue). The centroids of the three distributions are represented as gold stars, and their associated phylogenetic trees are visualized.

tumor (1), liver metastases ($\sim 5$), non-liver metastases ($\sim 5$).

All combinatorial 4-trees are inferred from this data and their mapping to $\mathbb{P}\Sigma_4$ is visualized in Figure 7. We denote the normal tissue sample by N, the primary pancreatic tumor by P, the liver metastases by LM, and the non-liver metastases by nLM. Contrary to our hypothesis that liver metastases give rise to metastases in other tissues, we find that the centroid of the data corresponds to a tree with branched ancestry between LM and nLM. Furthermore, we observe that there is no branching in the progression from normal tissue to primary disease to metastatic potential. In other words, the trajectory leading to the common ancestor of LM and nLM is a linear one.

## 4 Conclusions

The recent surge in high throughput sequencing of cancer provides a window into the molecular events underlying oncogenesis, tumor progression, acquired drug resistance, and metastasis. Longitudinal sampling of of an individual's disease course is becoming more common and allows us to interrogate the *evolutionary history* of a given tumor. Here we have proposed a framework in which to analyze and compare sets of cancer evolutionary histories through a natural application of the work on BHV spaces [4] to the setting of unrooted phylogenetic trees representing sequential tissue samples from an individual's disease. As the collection of longitudinal genomic data sets accelerates, it will become unfeasible to directly reason about large forests of phylogenetic trees. Our framework directly aids in the visual and statistical exploration of tumor evolutionary data with the goal of augmenting the intuition of cancer biologists and oncologists. Exploring the distributions of evolutionary modes in different cancers can have implications for personalized medical management and prognostication. For example, tuning the intensity of a patient's chemotherapeutic regimen requires an understanding of how treatment perturbs the natural evolutionary path of the cancer.

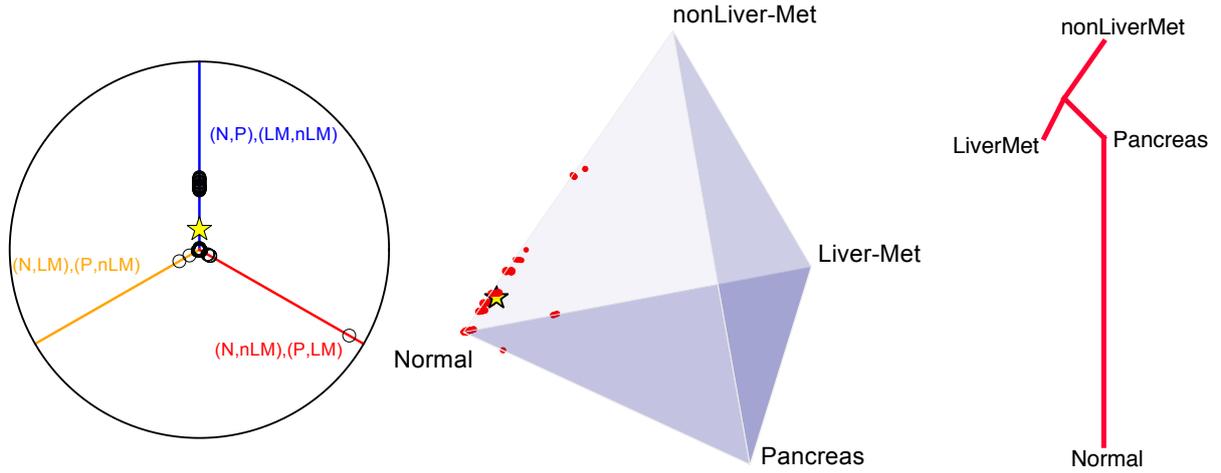

Figure 7: Both linear and branching behavior observed in 10 cases of metastatic PDAC. A strong tendency toward (N,P),(LM,nLM) topology in is seen in the star plot on the left. The majority of genetic alterations are acquired at the primary tumor stage. Evolution to LM and nLM do not appear to be linearly related. The centroid of the distribution is represented as a gold star, and its associated phylogenetic tree is visualized.

In this paper we have directly applied this framework to the analysis of three different cancers exhibiting distinct evolutionary patterns. In the case of acute myelogenous leukemia the highly localized distribution of patient histories in $\mathbb{P}\Sigma_3$ raises interesting questions about possible mechanisms of relapse–associated evolution. Follicular lymphoma evolves as a highly branched process, where the dominant clone in FL is not the direct ancestor of the dominant clone after transformation. Lastly, in the case of pancreatic ductal adenocarcinoma the metastases appear to derive linearly from the primary tumor, although the liver lesions and non–liver lesions branch from a common ancestor.

Our efforts are now turning toward developing statistical and machine learning methods for extracting meaningful biological and clinical information from point cloud data in $\Sigma_m$. Simulations of archetypal evolutionary scenarios are also being developed in the service of performing statistical inference on patient histories represented in our evolutionary moduli spaces [22]. One particular goal of ours is to characterize shifts in point clouds as a function of drugs administered to the patients, making use of clinically labeled data sets. From a patient management standpoint one would like to quantify distances between sets of treated and un-treated tumor histories and to look for correlations between regions of the space and clinical variables. The falling cost of sequencing will permit ever denser sampling of cancer evolution, and creativity will be needed if visualizations are to keep pace.

## Acknowledgments

The authors gratefully acknowledge the constructive feedback of Adolfo Ferrando, Teresa Palomero, Ricardo Dalla-Favera, Laura Pasqualucci, and Kevin Emmett. This work is supported by NIH grants R01 CA185486, R01 CA179044, and U54 CA121852, as well as the Stewart Foundation.

# References

[1] J. Aitchison. The Statistical Analysis of Compositional Data. *Journal of the Royal Statistical Society*, 44(2):139–177, 1982.

[2] A. Alexandrov. Über eine verallgemeinerung der riemannschen geometrie. *Schr. Forschungsinst. Math. Berlin*, 1:33–84, 1957.

[3] D. Barden, H. Le, and M. Owen. Central limit theorems for Fréchet means in the space of phylogenetic trees. *Electron. J. Probab.*, 18:1–25, 2013.

[4] L. J. Billera, S. P. Holmes, and K. Vogtmann. Geometry of the Space of Phylogenetic Trees. *Advances in Applied Mathematics*, 27(4):733–767, Nov. 2001.

[5] P. J. Campbell, S. Yachida, L. J. Mudie, P. J. Stephens, E. D. Pleasance, L. a. Stebbings, L. a. Morsberger, C. Latimer, S. McLaren, M.-L. Lin, D. J. McBride, I. Varela, S. a. Nik-Zainal, C. Leroy, M. Jia, A. Menzies, A. P. Butler, J. W. Teague, C. a. Griffin, J. Burton, H. Swerdlow, M. a. Quail, M. R. Stratton, C. Iacobuzio-Donahue, and P. A. Futreal. The patterns and dynamics of genomic instability in metastatic pancreatic cancer. *Nature*, 467(7319):1109–13, Oct. 2010.

[6] E. Carlotti, D. Wrench, J. Matthews, S. Iqbal, A. Davies, A. Norton, J. Hart, R. Lai, S. Montoto, J. G. Gribben, T. A. Lister, and J. Fitzgibbon. Transformation of follicular lymphoma to diffuse large B-cell lymphoma may occur by divergent evolution from a common progenitor cell or by direct evolution from the follicular lymphoma clone. *Blood*, 113(15):3553–7, Apr. 2009.

[7] L. Ding, T. J. Ley, D. E. Larson, C. a. Miller, D. C. Koboldt, J. S. Welch, J. K. Ritchey, M. a. Young, T. Lamprecht, M. D. McLellan, J. F. McMichael, J. W. Wallis, C. Lu, D. Shen, C. C. Harris, D. J. Dooling, R. S. Fulton, L. L. Fulton, K. Chen, H. Schmidt, J. Kalicki-Veizer, V. J. Magrini, L. Cook, S. D. McGrath, T. L. Vickery, M. C. Wendl, S. Heath, M. a. Watson, D. C. Link, M. H. Tomasson, W. D. Shannon, J. E. Payton, S. Kulkarni, P. Westervelt, M. J. Walter, T. a. Graubert, E. R. Mardis, R. K. Wilson, and J. F. DiPersio. Clonal evolution in relapsed acute myeloid leukaemia revealed by whole-genome sequencing. *Nature*, 481(7382):506–10, Jan. 2012.

[8] A. Feragen, M. Owen, J. Petersen, M. M. W. Wille, L. H. Thomsen, A. Dirksen, and M. de Bruijne. Tree-space statistics and approximations for large-scale analysis of anatomical trees. *Information processing in medical imaging : proceedings of the ... conference*, 23:74–85, Jan. 2013.

[9] M. Greaves and C. C. Maley. Clonal evolution in cancer. *Nature*, 481(7381):306–13, Jan. 2012.

[10] S. P. Holmes. Bootstrapping trees: theory and methods. *Statist. Sci.*, 18:241–255, 2003.

[11] S. P. Holmes. Statistics for phylogenetic trees. *Theoretical population biology*, 63:17–32, 2003.

[12] S. P. Holmes. Statistical approach to tests involving phylogenies. In O. Gascuel, editor, *Mathematics of evolution and phylogeny*. Oxford University Press, 2007.

[13] S. Li, T. Hricik, S. S. Chung, H. Bar, A. L. Brown, J. P. Patel, F. Rapoport, L. Liu, C. Sheridan, J. Ishii, P. Zumbo, J. Gandara, I. D. Lewis, L. B. To, M. W. Becker, M. L. Guzman, R. J. D'Andrea, F. Michor, C. Y. Park, M. Carroll, R. L. Levine, C. E. Mason, and A. M. Melnick. Epigenetic deregulation in relapsed acute myeloid leukemia. *Blood*, 122(21):2499, 2013.

[14] E. Miller, M. Owen, and J. S. Provan. Polyhedral computational geometry for averaging metric phylogenetic trees. 2014.

[15] P. C. Nowell. The clonal evolution of tumor cell populations. *Science (New York, N.Y.)*, 194(4260):23–8, Oct. 1976.

[16] S. Ohtake, S. Miyawaki, H. Fujita, H. Kiyoi, K. Shinagawa, N. Usui, H. Okumura, K. Miyamura, C. Nakaseko, Y. Miyazaki, A. Fujieda, T. Nagai, T. Yamane, M. Taniwaki, M. Takahashi, F. Yagasaki, Y. Kimura, N. Asou, H. Sakamaki, H. Handa, S. Honda, K. Ohnishi, T. Naoe, and R. Ohno. Randomized study of induction therapy comparing standard-dose idarubicin with high-dose daunorubicin in adult patients with previously untreated acute myeloid leukemia: the JALSG AML201 Study. *Blood*, 117(8):2358–65, Feb. 2011.

[17] J. Okosun, C. Bödör, J. Wang, S. Araf, C.-Y. Yang, C. Pan, S. Boller, D. Cittaro, M. Bozek, S. Iqbal, J. Matthews, D. Wrench, J. Marzec, K. Tawana, N. Popov, C. O'Riain, D. O'Shea, E. Carlotti, A. Davies, C. H. Lawrie, A. Matolcsy, M. Calaminici, A. Norton, R. J. Byers, C. Mein, E. Stupka, T. A. Lister, G. Lenz, S. Montoto, J. G. Gribben, Y. Fan, R. Grosschedl, C. Chelala, and J. Fitzgibbon. Integrated genomic analysis identifies recurrent mutations and evolution patterns driving the initiation and progression of follicular lymphoma. *Nature genetics*, 46(2):176–81, Feb. 2014.

[18] M. Owen and J. Provan. A fast algorithm for computing geodesic distances in tree space. *IEEE/ACM Transactions on Computational Biology and Bioinformatics*, 8(1):2–13, 2011.

[19] L. Pasqualucci, H. Khiabanian, M. Fangazio, M. Vasishtha, M. Messina, A. B. Holmes, P. Ouillette, V. Trifonov, D. Rossi, F. Tabbò, M. Ponzoni, A. Chadburn, V. V. Murty, G. Bhagat, G. Gaidano, G. Inghirami, S. N. Malek, R. Rabadan, and R. Dalla-Favera. Genetics of follicular lymphoma transformation. *Cell reports*, 6(1):130–40, Jan. 2014.

[20] K. Sturm. Probability measures on metric spaces of nonpositive curvature. *Contemporary mathematics*, 0000:1–34, 2003.

[21] G. Tzoneva, A. Perez-Garcia, Z. Carpenter, H. Khiabanian, V. Tosello, M. Allegretta, E. Paietta, J. Racevskis, J. M. Rowe, M. S. Tallman, M. Paganin, G. Basso, J. Hof, R. Kirschner-Schwabe, T. Palomero, R. Rabadan, and A. Ferrando. Activating mutations in the NT5C2 nucleotidase gene drive chemotherapy resistance in relapsed ALL. *Nature medicine*, 19(3):368–71, Mar. 2013.

[22] S. Zairis, H. Khiabanian, A. J. Blumberg, and R. Rabadan. Statistical learning in evolutionary moduli spaces. *To appear*.